\documentclass[superscriptaddress,secnumarabic,amssymb,amsmath,nobibnotesquantum universes,showpacs,aps,prd,nofootinbib,twocolumn]{revtex4}
\usepackage{color}
\usepackage{graphicx}
\usepackage{dcolumn}
\usepackage{bm}

\newcommand{\beq}{\begin{equation}}
\newcommand{\eeq}{\end{equation}}
\newcommand{\bey}{\begin{eqnarray}}
\newcommand{\eey}{\end{eqnarray}}

\begin{document}

\title{Geometrical Origin of Growth of Matter Perturbations}

\author{S S  De}
\email{ desatya06@gmail.com} \affiliation{Department of
Applied Mathematics, University of Calcutta, Kolkata 700009,
India}

\author{Farook Rahaman}
\email{farookrahaman@gmail.com} \affiliation{Department of
Mathematics, Jadavpur University, Kolkata 700032, West Bengal,
India}

\author{Antara Mapdar}
\email{antaramapdar@gmail.com} \affiliation{Department of
Mathematics, Shri Shikshayatan College,
11, Lord Sinha Road,  Kolkata 700071, West Bengal,
India}

\date{\today}

\begin{abstract}
 The density perturbation of the universe  has been considered in the framework of a Finslerian cosmological model in which the background spacetime of the universe is taken as the spatially flat FLRW spacetime with a Finslerian perturbation. The inhomogeneity in the matter(energy) density arises naturally in this consideration of gravity in the $( \alpha, \beta)$ - Finsler space for the background spacetime of the universe. The resulting inhomogeneous matter density indicates matter perturbation departing from the uniform density, and it is caused by the Finslerian perturbation in the Riemannian spacetime, thus ascertaining the geometrical origin of the growth of matter perturbation.\\
\\


\end{abstract}


\maketitle

\section{Introduction}
 For the galaxy formation and clustering in our universe, it is required to have relic fluctuations or the fractional density perturbation to be generated to some desired magnitude. The cosmological inhomogeneity departing from the uniform density   that causes the structure formation in the universe has been studied in the standard cosmological models on the basis of various processes that can lead to the gravitational instability necessary for such creation of fluctuations. These processes include the amplification of quantum zero-point fluctuations during inflation, the topological defects such as cosmic strings formed during the cosmological phase transition etc. Particularly, the linear growth dynamics of matter perturbations has been considered with the gravitational and hydrodynamical process with peculiar vector field which might be reluctant for structure formation. A good account of all these studies under standard models based on Riemannian geometry can be found in  \textcolor{blue}{[1-3]}. Of course, growth of density perturbation has been made in the Finsler-Rander cosmology \textcolor{blue}{[4]}. Finsler geometry is, in fact, a generalization of Riemannian geometry, that naturally generalizes the gravitational field equations of general relativity. So far, in these approaches of the linear growth of perturbation, the processes involving cold dark matter, collision matter and scalar field have to be introduced. Recently , we have introduced a Finslerian cosmology\textcolor{blue}{[5, 6]} and here we shall show how this consideration can account for the density perturbation without any additional input into the system, thus, making it possible to ascertain the geometrical origin of inhomogeneity and anisotropy.  \\
The paper is organized as follows. In section 2, a brief introduction of our cosmological model is given. With the barotropic equation of state, the equation for scale factor has been derived from the gravitational  field equations of the Finsler spacetime considered in earlier work \textcolor{blue}{[6]}. In section 3, we find growth rate and growth index of the density perturbation from the solution obtained for the scale factor. In the final section 4, some concluding  remarks have been made. We shall use Planck units $\hbar = G = C=1$

 \section{Gravitational Field Equations}

$~~~~~~$ For the background spacetime of the universe we have introduced the following Finslerian Structure \textcolor{blue}{[5,6]}:
 \begin{equation}
     F^2=y^ty^t-a^2(t)y^ry^r- r^2a^2(t)\bar F^2(\theta, \phi, y^\theta, y^\phi),
 \end{equation}
 where $\bar F^2 $ is regarded as the Finsler structure of the two-dimensional Finsler space. $\bar F^2$ was proposed to be of the form:

 \begin{equation}
 \bar{F^2} = y^\theta y^\theta+f(\theta,\phi)y^\phi y^\phi.
 \end{equation}
  In Finsler geometry, there is a geometrically invariant Ricci scalar $Ric \equiv R^\mu_\mu$. It depends only on the Finsler structure and is insensitive to connections , such as Chern connections, Cartan connection etc. \\
 $~~~~~~~$ Now if the function $f$ is independent of $\phi$ , that is,      $ f(\theta, \phi) = f(\theta)$, then the Ricci scalar $\bar {Ric}$ for the Finsler structure $\bar F$
 is given by
 \begin{equation}
      \bar{Ric} = - \frac{1}{2f}\frac{d^2f}{d\theta^2}+ \frac{1}{4f^2}\left (\frac{df}{d\theta}\right )^2.
      \end{equation}
 For constant or $\theta$-dependent flag curvature, that is for $\bar{Ric}=\lambda(\theta)$, we have the following equation for   specification of the function $f(\theta)$:

  \begin{equation}
       - \frac{1}{2f}\frac{d^2f}{d\theta^2}+ \frac{1}{4f^2}\left (\frac{df}{d\theta}\right )^2=\lambda(\theta).
  \end{equation}
  In the appendix A, we have discussed the solutions of this equation. The modified gravitational field equations can be found with the general energy-momentum tensor for matter distribution given as
  \begin{equation}
      T^{\mu}_{\nu}=(\rho+p_t)u^\mu u_\nu-p_tg^\mu_\nu+(p_r-p_t)\eta^\mu \eta_\nu,
  \end{equation}
  where $u^\mu u_\nu=-\eta^\mu \eta_\nu=1$; $p_r$, $p_t$ being respectively the radial and transverse pressures for the anisotropic fluid. These are

  \begin{equation}
      8\pi_F G\rho=\frac{3 \dot a^2}{a^2}+\frac{\lambda-1}{r^2a^2}
  \end{equation}
  \begin{equation}
      8\pi_F Gp_r=-\frac{2\ddot a}{a}-\frac{ \dot a^2}{a^2}-\frac{\lambda-1}{r^2a^2},
  \end{equation}
  \begin{equation}
      8\pi_F Gp_t=-\frac{2\ddot a}{a}-\frac{ \dot a^2}{a^2}.
  \end{equation}
  As in \textcolor{blue}{[5]}, we consider the barotropic equation of state
  \begin{equation}
  P=\omega\rho
  \end{equation}
  where the pressure P is given by
  \begin{equation}
      P=(1+\omega)p_t-\omega p_r-\frac{m^2r^3}{2}F_a.
  \end{equation}
  Here $F_a$ is the anisotropic force which is
  \begin{equation}
      F_a=\frac{2(p_t-p_r)}{r}.
  \end{equation}
  Then the equation foe scale factor $a(t)$ follows. It is given as
 \begin{equation}
     \ddot a+\frac{1+3\omega}{2}\frac{\dot a^2}{a} -\frac{m^2(\lambda-1)}{2a}=0.
 \end{equation}
 It is to be noted that the background Finslerian spacetime of the universe, that we are considering here is, in fact, a spatially flat FLRW spacetime of scale factor $a(t)$ with Finslerian perturbation as shown in\textcolor{blue}{ [5,6]}.

 \section{Growth of Density Perturbation}
 We write the equation (6) in the following form:
 \begin{equation}
    \rho(t,r,\theta)=\frac{3\dot a^2}{8\pi_F Ga^2}+\delta\rho(t,r,\theta),
     \end{equation}
 where\begin{equation}
     \delta\rho=\frac{\lambda-1}{8\pi_F G r^2a^2}.
 \end{equation}
   Then the density perturbation $\delta_m$ is $\frac{\delta\rho}{\rho_c}$, where
 \[\rho_c=\frac{3\dot a^2}{8\pi_F Ga^2} \equiv \frac{3H^2}{8\pi_FG},\]
 i.e., \begin{equation}
     \delta_m=\frac{\epsilon\phi(\theta)}{3r^2\dot a^2}.
 \end{equation}
   Where $\lambda= 1+\epsilon \phi(\theta)$ with small $\epsilon$ (see appendix A)\\
  For growth rate of clustering we use the following relation \textcolor{blue}{[3,4]}:
  \begin{equation}
      f(a)=\frac{d(\ln\delta_m)}{d(\ln a)} \simeq\left[\Omega_m(a)\right]^\gamma,
  \end{equation}
  where $\gamma$ is the growth index and $\Omega_m(a)$ is given by

  \begin{equation}
      \Omega_m(a)=\frac{\rho(t,r,\theta)}{\rho_c}=1+\frac{\epsilon\phi(\theta)}{3\dot r^2\dot a^2} =1+\delta_m.
  \end{equation}
   Now by using (16) \begin{equation}
      f(a)=\frac{a\dot \delta_m}{\dot a \delta_m} =-\frac{2a\ddot a}{\dot a^2}=q,
  \end{equation}
  where $q$ is the deceleration parameter. Therefore we have from (16)
  \begin{equation}
      q=(1+\delta_m)^\gamma.
  \end{equation}
  Also, by using (12), we have from (19)
  \begin{equation}
            q=1+3\omega-\frac{\epsilon m^2\phi(\theta)}{\dot a^2}=(1+\delta_m)^\gamma.
  \end{equation}
  The pressure $P$, which is given in (10) can be written as
  \begin{equation}
      P=(1+\omega)p_t-\omega p_r-P_a,
  \end{equation}
  where the anisotropic pressure $P_a$ is given by
  \begin{equation}
      P_a=k\frac{r}{2}F_a=k(p_t-p_r),     ~~ [ using !~(11) ]
  \end{equation}
  in which $m$ has been specified as \\
 $~~~~~~~~~~~~~~~~~~~~~~$ $m^2r^2=k$ ,  ($k$ is a small function of $r$)

\begin{figure*}
\includegraphics[scale=0.3]{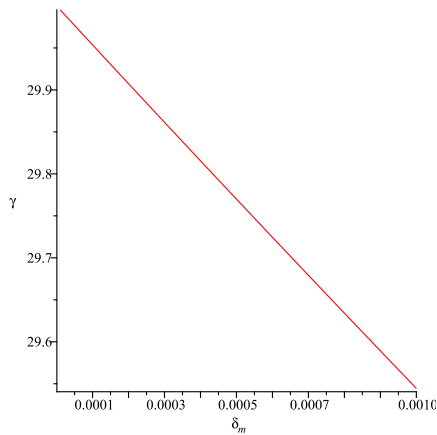}
\caption{Variation of growth index with respect to  $\delta_m$.}
\label{fig5}
\end{figure*}
 Consequently, we have from (22)
 \begin{equation}
     P=(1+\omega-k)p_t-(\omega-k)p_r.
 \end{equation}
  Thus, this pressures is the weighted average of the radial and transverse pressures of the anisotropic fluid.\\
  From the equations (7) and (8), we have

  \begin{equation}
      p_t-p_r=\frac{\lambda-1}{8\pi_F G r^2a^2} =\frac{\epsilon\phi(\theta)}{8\pi_F Gr^2a^2}=\delta \rho.
  \end{equation} ~ [By using (14)]\\
  Also, from (22), we have\\
  $~~~~~~~~~~~~~~~~~~~~~~~~~~~~~$ $P_a= k\delta \rho$, [by using (24)]\\
  or, \begin{equation}
      \frac{P_a}{\rho_c}=k\frac{\delta\rho}{\rho_c}=k\delta_m.
  \end{equation}
  From (20), we have

  \begin{equation}
      (1+\delta_m)^\gamma=1+3\omega-\frac{\epsilon k\phi(\theta)}{ r^2 \dot a^2}=1+3\omega-3k\delta_m.
  \end{equation}
  $~~~~~~~~~~~~~~~~~~~~~~~~~~~~~~~~~~~~~~~~~~~~~ ~~~~~~~~~~~[By ~using ~(15)]$

  Then, the growth index is given as
  \begin{equation}
      \gamma= \frac{\ln\{1+3(\omega-k\delta_m)\}}{\ln(1+\delta_m)}.
  \end{equation}
  Now, we find the solution of the equation (12) for the scale factor $a(t)$ for the matter-dominated era with small pressure $P$, that is, for $P=p_t=k\rho$ by taking $\omega=k$. In this case we find the solution to be
  \begin{equation}
      a(t)=\left(\frac{t}{t_0}\right)^{\frac{2}{3}(1-k)},
  \end{equation}
  $a(t)$ being normalized to 1 at the present epoch $t_0$. On the other hand, the solution for radiation era , that is for $\omega=\frac{1}{3}$ or the pressure $P=\frac\rho{3}$ can be found to be
  \begin{equation}
      a(t) \propto t^{\frac{1}{2}}.
  \end{equation}
  Then, for matter-dominated era, density perturbation is \\
  $~~~~~~~~~~~~~~~~~~~~~~~~~~$ $\delta_m=\frac{3\epsilon \phi(\theta)}{4(1-k)^2} \left(\frac{t_0}{r}\right)^2 \left(\frac{t}{t_0}\right)^{2k} a(t)$,
  that is, \begin{equation}
      \delta_m \propto \left(\frac{t}{t_0}\right)^{2k}a(t).
  \end{equation}
  For radiation era, the density perturbation is proportional to $a^2(t)$.\\
  The growth index for the matter-dominated era is, from equation (27),
  \begin{equation}
      \gamma= \frac{\ln\{1+3k(1-\delta_m)\}}{\ln(1+\delta_m)},
  \end{equation}
   and the growth index for the radiation era is
  \begin{equation}
      \gamma= \frac{\ln(2-3k\delta_m)}{\ln(1+\delta_m)}.
  \end{equation}
  The density perturbation for the important case of matter-dominated era is given in (30), and there the constant of proportionality is $O(k)$. For the present epoch $t_0$ we then have $\delta_m$ is of the order of $k$. Consequently, we are giving here a plot of the growth index for the different values of $\delta_m$, that is for $\delta_m= 10^{-5}$ to $10^{-3}$ (Fig 1)

  \section{Concluding Remarks}
  We have considered density perturbation of a Finslerian cosmological model in which the background spacetime is FLRW spacetime with Finslerian perturbation. This background spacetime of the universe has been shown \textcolor{blue}{[5]} to be a $(\alpha, \beta)$ - Finsler space. By using the barotropic equation of state for the anisotropic fluid , we have derived the equation for scale factor $a(t)$ from the modified gravitational field equations of the Finsler space, which is, in fact, a natural generalisation  of Riemann space . It has been possible here to identify the inhomogeneous portion of the matter density as the density perturbation departing from its uniform part. Also, from the solutions of the scale factor, the growth indices for the growth of density perturbations necessary for structure formation in the universe have been obtained for both matter-dominated and radiation eras. Here, we should note that the inhomogeneity in density does not appear in the case $\lambda=1$, which corresponds to spatially flat FLRW spacetime(the Riemannian spacetime) without any Finslerian perturbation. Thus, this Finslerian perturbation is responsible for the density perturbation. The magnitude of density perturbation  depends on the order of smallness in the difference of
  $\lambda$ from unity. We, thus, find the geometrical origin of the density perturbation without introducing any classical and quantum ingredients into the system for producing such gravitational instabilities.

\vskip .5cm
 \noindent{\textbf{APPENDIX A}}
\vskip .5cm
In finding solution of the equation
\begin{equation}\tag{A1}
           -\frac{1}{2f}\frac{d^2f}{d\theta^2}+ \frac{1}{4f^2}\left (\frac{df}{d\theta}\right )^2=\lambda(\theta)=1+\epsilon\phi(\theta),
\end{equation}
we write , \begin{equation}\tag{A2}
    \frac{1}{f} \frac{df}{d\theta}=F(\theta)
\end{equation}
Then, we have
\begin{equation}\tag{A3}
    \frac{dF}{d\theta}+\frac{1}{2}F^2=-2\lambda(\theta)=-2(1+\epsilon\phi(\theta)) \simeq-2e^{\epsilon\phi(\theta)}.
\end{equation}
Now, the two dimensional Finslerian structure $\bar  F^2(\theta, \phi, y^\theta, y^\phi)$ has been chosen as
\begin{equation}\tag{A4}
    \bar F^2= y^\theta y^\theta+f(\theta)y^\phi y^\phi.
\end{equation}
  Setting \begin{equation}\tag{A5}
     f(\theta)=\sin^2\theta+\epsilon \chi(\theta)
 \end{equation}
  we have
 \begin{equation}\tag{A6}
     \begin{split}
F(\theta)=\frac{1}{f} \frac{df}{d\theta}= \frac{d}{d\theta}\ln \left[\sin^2\theta \left(\frac{\epsilon\chi(\theta)}{\sin^2\theta}\right)\right] \\ \simeq \frac{d}{d\theta}\ln \left[(\sin^2\theta) e^{\frac{\epsilon\chi(\theta)}{\sin^2\theta}}\right]\\
     = 2\cot \theta+ \epsilon \frac{d}{d\theta}\left(\frac{\chi(\theta)}{\sin^2\theta}\right)
      \\= 2 \cot \theta + \epsilon \bar \chi(\theta),
      \end{split}
   \end{equation}
  where, \begin{equation}\tag{A7}
      \bar \chi(\theta)=\frac{d}{d\theta}\{\frac{  \chi(\theta)}{\sin^2\theta}\}.
  \end{equation}
  Then, \begin{equation}\tag{A8}
      \frac{dF}{d\theta}=-\frac{2}{\sin^2\theta}+\epsilon \bar \chi '(\theta).
  \end{equation}
  From (A3) and (A8), we have \\
  \[-\frac{2}{\sin^2\theta}+\epsilon \bar \chi '(\theta)+ \frac{1}{2}\{4\cot^2\theta+4\epsilon \bar\chi(\theta)\cot \theta+\epsilon^2 \bar\chi(\theta)\} \]\[ =-2(1+\epsilon \phi(\theta)),\]
  or,
  \[-\frac{2}{\sin^2\theta}(1-\cos^2\theta)+\epsilon\{ \bar \chi '(\theta)+ 2 \bar\chi(\theta)\cot \theta\}
  =-2-2\epsilon\phi(\theta)).\]
  $~~~~~~~~~~~~~~~~~~~~~~~~~~~~~~~~~~~~~~~~~~~~~~~$[neglecting $\epsilon^2$- term]\\
  Therefore, we must have 
 \[~~~~\bar \chi'(\theta)+2\bar\chi(\theta)\cot \theta +2\phi(\theta)=0,\]
  or,
\[~ \frac{d}{d\theta}\{\sin^2 \theta\bar \chi(\theta)\}= -2\sin^2 \theta \phi(\theta).\]
  On integration we have
  \begin{equation}\tag{A9}
      \bar\chi(\theta)=-\frac{2}{\sin^2\theta}\int \sin^2\theta \phi(\theta) d\theta +\frac{A}{\sin^2\theta},
  \end{equation} $~~~~~~~~~~~~~~~~~~$ where A is the integrating constant.\\
  Using (A7), we have from (A9)
  \begin{equation}\tag{A10}
      \frac{d}{d\theta}\{\frac{\chi(\theta)}{\sin^2\theta}\}=-\frac{2}{\sin^2\theta}\int \sin^2\theta \phi(\theta) d\theta +\frac{A}{\sin^2\theta}.
  \end{equation}
  Now, we consider some useful cases. \\

  Case 1: Let $\phi(\theta)=\cos\theta$:\\
  
  In this case we get from (A10),
 \[\frac{d}{d\theta}\{\frac{\chi(\theta)}{\sin^2\theta}\}=-\frac{2}{3}\sin\theta +\frac{A}{\sin^2\theta}.\]
  Therefore, on integration, we have \\
   \[~~~~\chi(\theta)=\frac{2}{3}\sin^2\theta \cos \theta -A\cos\theta\sin\theta+B\sin^2\theta,\] where $B$ is an integrating constant.
   \[~~ If~A = B =0 ,~ \chi(\theta)=\frac{2}{3}\sin^2\theta \cos \theta.\]
  Consequently, we have \\
  \[~ f(\theta)=\sin^2\theta\left(1+\frac{2\epsilon}{3}\cos\theta\right),\]
  and
\[~~~~~\phi(\theta)=\cos\theta.\] [by using (A1)]\\

  Case 2:   $\chi(\theta)=\sin^2\theta(A_1+A_2\theta+A_3\theta^2)$:\\
  
 Here we can find  
 \[~~~~\phi(\theta)=A_3-\cot\theta(A_2+2A_3\theta).\] 
 
  Case 3:  $\chi(\theta)=(\sin^2\theta)e^{-A\theta^2}$:\\
  
   Here we can find
  \[~~\phi(\theta)=(A-2A^2\theta^2+A\theta\cot\theta)e^{-A\theta^2}.\]

 It will be interesting to see the observational result on the inhomogeneity in the matter distribution that can ascertain $\phi(\theta)$.

\section*{Acknowledgments}

FR would like to thank the authorities of the Inter-University Centre for Astronomy and Astrophysics, Pune, India for providing research facilities.

{}

\end{document}